# The ballistic method of pumping solid-state lasers

Volov Dm.
Samara State Transport University
Samara, Russia
volovdm@mail.ru

Shmelev V.M.
Institute of Chemical Physics RAS
Moscow, Russia

**The ballistic method for obtaining a dense radiating plasma and the possibilities of using this method to pump solid-state lasers are investigated. The plasma was obtained experimentally by heating the working gas in a ballistic plasmatron. A vortex chamber is proposed for transferring energy into the plasmatron - laser system.**

*Optical pumping of lasers; a ballistic plasmatron; dense radiating plasma; a vortex chamber*

## I. Introduction

Powerful sources of visible and ultraviolet radiation are required in order to accomplish diverse photochemical transformations and to solve a number of technological problems, for optical pumping of lasers [1]. At the present time, pulsed gas-discharge tubes have found wide application for these purposes. The source of radiation is an electric-discharge plasma with a temperature of 10 to 12 thousand degrees.

An alternative, nonelectrical method for obtaining a radiating plasma could be heating the working gas up to temperatures 6 to 12 thousand degrees with adiabatic compression of the gas in a special apparatus – a ballistic plasmatron. However, experiments [2] on single adiabatic compression of a gas by a freely flying piston showed the ballistic plasmatron have a low energy efficiency as a source of optical radiation as compared with pulsed gas tubes.

An effective method for increasing the specific energy of a plasmatron is preheating the working gas [3]. This can be achieved by using several stages of compression and expansion using intermediate membranes, valves, and pistons in the system.

## II. The Ballistic Pumping in Multistage Compressed Heated Gas

A common driver technique for shock tubes is the free-piston compressor as described by Stalker (1972) [4]. These compressors have proven to be versatile, however, driver gas pressure and temperature are determined (and limited) by the volumetric compression ratio. This means that to achieve the temperature required for high speed shock tube applications, large compression ratios of 1/100 are needed. One alternative is to employ compression processes, which include an entropy-raising stage that can produce a given temperature rise with a lower compression ratio than an isentropic process. This means that a shock tube employing such processes can be built smaller (and more cheaply). Two mechanisms available for raising the driver gas entropy are shock heating and throttling. A preliminary experimental study was carried out on a small scale shock-assisted free-piston driver which stemmed from a concept suggested by Bogdanoff (1990) [5].

The concept coupled quasi-isentropic free-piston compression with shock waves generated by rupture of a diaphragm placed within the driver. Temperature increases of up to 50% (Kendall (1997) [6]) over a standard free-piston compressor were yielded with an argon driver gas, while only marginal gains were achieved with helium. These final driver gas temperatures, however, fell well below ideal predictions, due to heat losses, particularly behind the incident and reflected shocks.

Throttling processes have been combined successfully with free-piston compressors to raise the driver gas entropy and performance. Shmelev (1993) [7] applied a dual-piston technique where a heavier piston compressed the driver gas that throttled through the hole of a lighter piston placed within the driver. Operation of the concept, however, was limited in practice by the complexity of tuning both piston trajectories with pistons of varying masses and position. Similarly, in the PGU units in Russia (Anfimov (1992) [3]), free piston compression was combined with the throttling of the driver gas and controlled with complex specialized high speed valves to enhance the final temperatures achieved in the driver.

Various methods of gas compression are reviewed, which are intended for heating the gas. Special attention is given to units of nonisentropic compression of the mechanical heat-engineering type, in which the gas is transferred from chamber to chamber in order to increase the temperature [1].

The problem of radiation energy transfer in ballistic radiative-gas compression facilities is considered. For constructing mathematical models of these facilities, an algorithm for splitting the thermodynamic system into sections (with consideration of the interconnections of the polysectional system) is used. The conditions of applicability of the thermodynamic approach to describing the processes in a ballistic plasmotron are established. It is shown that, in most practical problems, the losses in the intersectional channels are insignificant as compared with the losses in the sections and can be neglected [8]. Methods are discussed of mathematical description of processes occurring in units of this type, including analytical investigation of simplified qualitative models and numerical simulation in more complex cases close to reality.

In the present paper we propose a nonelectrical method for obtaining plasma with temperature up to 15000 K, using nonisentropic heating of the working gas in a ballistic apparatus with two-stage compression.

### III. Experimental Researches of the Vortex Chamber

A large increase in the efficiency of heating the working gas for laser pumping can be achieved in relatively simple, two-stage compression setups – ballistic plasmatrons with two freely moving pistons, combined with vortex chamber [9].

In this paper data of the experiment with the vortex chamber, whose geometric sizes correspond to parameters used at numerical calculations are cited. In such vortex chamber [9] the laser optical system can be used. The vibration of the ballistic unit is not transmitted to the optical system: these units are divided by a gap between quartz glass and a laser rod, completely isolated from mechanical influence.

In the experiment were used xenon, argon, helium, air and their mixes as working gases. The thrusting gas was air in all cases. The pressure of thrusting gas varied from 30 up to 200 atm. For obtaining these pressure atmosphere storage tanks were originally used, the compressor was applied afterwards.

Pressure measurements in the chamber were made with piezo-sensors, such as T6000, LX-610 and LX-604. The measurements of the energy of radiation in the chamber were making with the calorimeter IMO-2, the pulse profile – with the photosensor.

The general view of the installation is represented in fig. 1. The filing of pressure variation in a pulse was made at four points (fig. 2): 1 - at a chamber inlet; 2 - at a butt-end of the chamber with radius $r = 45$ mm, $\varphi = 50^0$; 3 - in the chamber opposite the inlet; 4 - on a symmetrical axis near to a butt-end and at the inlet contrary.

At the first point measurements were fixed with the sensor T6000 and measurement plate A1 (fig. 2). For measurements at point 2 at a butt-end of the chamber the sensor LX-610 and second measurement plate A2 were used. At the third point the measurements were fixed with the sensor LX-604 and were registered by the digital oscillograph C. A signal from the sensor there passes through a common-drain amplifier to an oscillograph. In the experiment with the perspex windows, where the measurements at point 4 were made, the sensor LX-604 was connected to the plate A2.

In general, the pressure increase at the first checkpoint with the increase of the energy of thrusting gas is consistent with the theoretical model. The significant deviations (up to 25%) are observed only for air as working gas.

On pressure in the barrel graphics significant (up to 20% from a level of a signal) acoustic noise created by the easy piston movement causing a shock wave in the barrel is detected.

When using a vortex chamber as the optical unit for pumping lasers most important are the values of pressure at the butt-end (point 2). These pressures are identical at both butt-ends. This is the point where gas slows down and from which the radiation process is spread. The pressure at butt-ends characterizes the intensity of radiation processes.

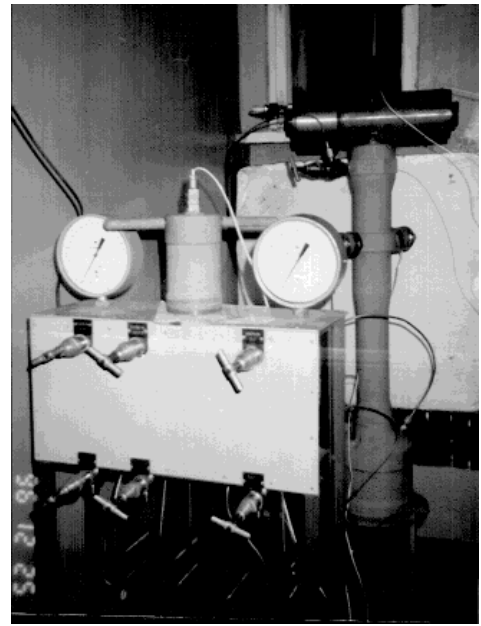

Figure 1. A general view of the installation.

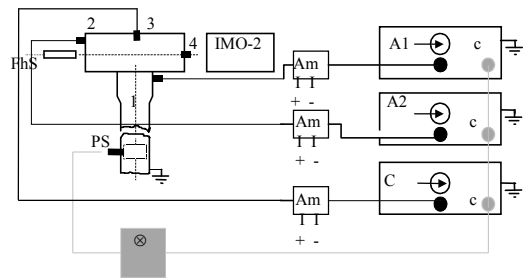

Figure 2. The scheme of measurements and start-up synchronization. PhS - photosensor; PS - position sensor; Am - common-drain amplifier; A1, A2 - measurement plates, C - digital oscillograph; c - synchronization jack.

The range of pressure changes at point 2 $p_2$ with the increase of energy is rather wide. The pressure at the chamber $p_1 = 300$atm (point 1) corresponds about 100atm at a butt-end, and with the increase of $p_1$ up to 500atm $p_2$ it was about 180atm. At point 2 there is a good conformity of maximum pressure in numerical and physical experiments (with a computational mode difference was not more than 10%).

The duration of oscillations reached 60ms. In all experiments the tendency of grouping a shock wave in series was registered. As was found out from a numerical experiment, these series occur due to the rotation of gas. However not one but serial are formed, which can't be found in a numerical experiment. The sequence of series with high pressures forms a characteristic picture, the so-called «crown».

To study the influence of the diameter of bulkhead section of a channel on the processes inside the chamber, and also with the purpose of finding the optimum cross-section area of a channel, before the inlet of the chamber the nozzles of various diameter were installed. These were spacers with minor diameters 6, 7.5, 8.5, 8.7, 13 mm with thick 1-2 mm. The installation of such nozzles enabled regulations of gas flow rate in the channel. The diameter of the nozzle also influences the size of the working gas compression in the barrel. The research was conducted with xenon with the pressure of thrusting gas 162.5 atm. Up to a certain moment the dependence of pressure at a butt-end on $d_c$ is practically linear, but at $d_c > 8.7$ mm the increase of pressure ceases, and at $d_c = 13$ mm a failure happens so it is impossible to achieve the pressure with large diameters of the nozzle. The analytical calculation gives $d_c = 9.4$ mm, which is 8% above the experimental. The limiting boundary $d_c$ appears rather sharp and the correct selection of the cross-section area of the nozzle in many respects determines the activity of the installation as a whole.

With other things being equal pressure graphics for different $d_c$ are completely combined on the time scale. The groups of waves arise with the intervals of $4 \pm 0.6$ ms.

When using the nozzle $d_c = 6$ mm the shock waves at the butt-end are practically not seen, in the chamber only the general increase of the pressure signal on all sensors is fixed. At $d_c = 6$ mm most of the radiating gas mass remains in the barrel during the piston reverse direction, without strong shock wave in the chamber. At $d_c = 7.5$ mm at the butt-end of the chamber the picture of shock waves ("crown"), inherent to the given process, is observed. The maximum pressure at the second point is 74 atm. Thus the second peak of pressure in the barrel is not so big. The maximum chamber pressures are observed at $d_c = 8.7$mm: 130atm, and in the barrel practically one peak of pressure appears. It is important to note, that the shape of pressure impulses at the butt-end of the vortex chamber observed at $d_c = 7.5$mm remains identical for all large $d_c$, only their intensity varies.

In table 1 data about the ratio of upstream pressures in vortex chamber and maximum pressure, occurring in this case, at the butt-end at $d_c = 8.7$mm are given. As it shows, the deviations from monotone increase of $p_2$ are observed. It is probably connected with the displacement of the point of maximum pressure from a checkpoint (phenomenon observed in numerical experiment) and with errors in pressure detecting.

Much greater intensity of shock waves in the chamber can be achieved using additional devices: a diaphragm 0.5 mm. So, the chamber pressures will increase up to 186 atm. The use of diaphragms results in a pressure grows in the barrel at the first pulse and does not influence the formation of the second pulse. Thus there is not need to search optimum the diameter for the nozzle cross-section.

At the third checkpoint the smooth pressure increase is observed.

TABLE I. THE EXPERIMENTAL DATA ON PRESSURE IN POINTS 1, 2

| N | $p_0$, atm | Point 1 | | Point 2 | |
|---|---|---|---|---|---|
| | | $p_1$, atm | sensor | $p_2$, atm | sensor |
| 1 | 75 | 242 | T6000-1 | 80 | LX -610 |
| 2 | 97.5 | 289 | T6000-1 | 98 | LX -610 |
| 3 | 100 | 300 | T6000-1 | 92 | LX -610 |
| 4 | 105 | 309 | T6000-3 | 111 | LX -610 |
| 5 | 125 | 331 | T6000-1 | 111 | LX -610 |
| 6 | 150 | 390 | T6000-3 | 127 | LX -610 |
| 7 | 157 | 403 | T6000-1 | 123 | T6000-1 |
| 8 | 162 | 428 | T6000-1 | 130 | LX -610 |
| 9 | 165 | 475 | T6000-1 | 139 | LX -610 |
| 10 | 177 | 544 | T6000-1 | 161 | T6000-1 |

As to the correspondence with the numerical experiment, at the computational mode $p_3$ it appears 10% higher. It can be explained by a wall roughness neglected in calculations and additional disturbances coming from relaxed waves reaching the sensor. The comparison of average $p_3$ is consistent with the experimental data - 47 atm in calculation and 42 atm on a computational mode.

Time of achieving the average pressure – is the time of chamber filling with parameters equalization: about 10 ms – corresponds to more precisely numerically designed (7.7 ms). Pressure at the third point at $p_0 = 180$ atm is 116 atm (copper diaphragm 0.5 mm thick, $d_c = 13$ mm is used).

Average chamber pressure in all experiments did not exceed 50 atm (according to the pressure measurement data at the third checkpoint).

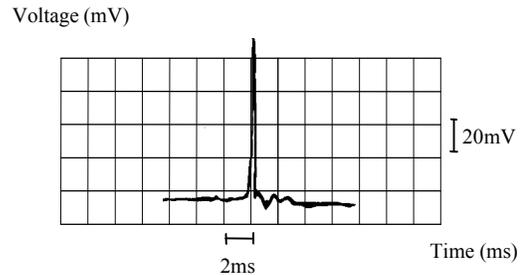

Figure 3. An impulse of radiation in the chamber.

In the experiments with the perspex a light flash both on Xe, and on an air was observed. The intensity radiation shape impulse always had one maximum (fig. 3). The duration did not exceed 1ms. Maximum radiation density in an impulse fixed by fhotosensor is $0.9 \cdot 10^{10}$ W/m$^3$. In view of the data of measurements IMO-2 it is possible to estimate full radiation energy in an impulse as $1 - 4$ kJ. Whereas the gas in the chamber remains as optical thick body, the radiation spectrum is close to the spectrum of an absolutely black body [8].

The comparison of the intensity of waves at a butt-end on an axis and rims shows, that at qualitative preservation of "crown" the level of pressure on the axis (44 atm) is 4... 5 times lower. It testifies that there a significant discharge of the central part in vortex version of the chamber. However it is necessary to note, that with the absence of the central body the circumferential pressure is lower. With xenon a shock wave in the second point is almost 1.5... 2 times weaker.

Main moment to confirm the efficiency of the central body unloading in the vortex chamber version was the test of the installation in the assembly with a quartz tube (diameter 40 mm). The experiments were made with xenon at the computational operational mode with maximum pressure of thrusting gas allowed for the given installation. The thickness of the quartz glass – 3 mm – was twice smaller, than in non-vortex versions of the system. At the previous installations it was not possible to achieve regular positive results and the glass 6 mm thick collapsed.

The experiments showed that when using the vortex chamber the glass even twice smaller thickness with other things being equal does not collapse, in difference from other types of chambers. Thus the efficiency of vortex chambers application in the installations of a similar type is experimentally proved.

As shown in other experiments [7], in two-phasic setup compression the energy of radiation of the plasma, surpassing the energy of radiation in one-phasic plasmatron and equal 5 – 18 $J/cm^2$ in a visible spectrum is reached, and it's 1.5 – 2 times more at a full spectrum of radiation. It is 5 – 10 times more surpass the specific energy of radiation one-phasic plasmatron. The best results are achieved in setup with the external valve, it is the most reliable and simple system. The reached transformation efficiency of gas energy to the light was ~20%, and in the visible spectrum of radiation ~10%.

## IV. Conclusions

The drop of impact loadings on the central body in vortex chambers is experimentally justified. In the considered installation the pressure on the central body does not exceed 100 atm, while in linear chambers these pressure achieve an order of 200 – 300 atm.

The results of the experiments on pressure measurement, the power and radiation pulse duration are consistent with the numerical calculation at checkpoints. The experimental research has shown, that in a vortex chamber because of redistribution of pressure on cross-section at butt-ends, near the axis area the pressure is essential (in 2 – 4 times) echanical influence on optical elements.